\documentclass[10pt,twocolumn]{revtex4}
\newcommand{\bR}{{\bf R}}

\usepackage{amsmath}
\usepackage{graphicx}
\usepackage{textcomp}
\begin{document}
\title{Types of single particle symmetry breaking in transition metal oxides due to electron correlation}
\author{Lucas K. Wagner}
\email{lkwagner@illinois.edu}
\affiliation{Dept. of Physics, University of Illinois at Urbana-Champaign}
\begin{abstract}
Very accurate wave functions are calculated for small transition metal oxide molecules.
These wave functions are decomposed using reduced density matrices to study the underlying  correlation of electrons.
The correlation is primarily of left-right type between the transition metals and the oxygen atoms, which is mediated by excitations from the nominal single Slater ground state into antibonding and d-type orbitals.
In a localized representation, this correlation manifests itself in a 2-electron hopping term that is off-diagonal.  
This term is of similar magnitude to the commonly considered Hubbard-type on-site interaction.
\end{abstract}
\maketitle

One of the grand challenges in modern condensed matter theory is the description and prediction of the properties of correlated electrons.  
Of particular interest are transition metal oxides, which exhibit effects including high T$_c$ superconductivity\cite{muller_discovery_1987,anderson_resonating_1987}, Mott insulator behavior\cite{mott_discussion_1937}, and colossal magnetoresistance\cite{basov_electrodynamics_2011,ramirez_colossal_1997}, all of which owe their existence to electron correlation effects.
Control of these strong correlation effects has the potential to open up many new areas in both physics research and technology, much like the control of weakly correlated electronic structure has enabled innumerable advances in the past 80 years since the development of that theory.
Much of the research in physics to date has concentrated on the development of phenomenological models of strong correlation, such as the  Hubbard model, that have had many successes in helping to understand these systems.
However, when considered from first principles, the underlying Coulomb Hamiltonian of a strongly correlated system is the same as for a weakly correlated system.
The difference between the two is an emergent property of the many-electron wave function.

There has been a large amount of effort devoted to treating strongly correlated systems starting from the first principles Hamiltonian.
These efforts have ranged from phenomenological corrections to density functional theory (DFT), such as DFT+U\cite{anisimov_band_1991,kulik_systematic_2010}/DFT+DMFT\cite{ren_lda+dmft_2006}, to other extensions of DFT using hybrid functionals, to GW perturbation theory\cite{jiang_first-principles_2010}, quantum chemistry\cite{jiang_multireference_2011}, and quantum Monte Carlo\cite{wagner_energetics_2007,kolorenc_wave_2010,al-saidi_auxiliary-field_2006,kolorenc_quantum_2008,foulkes_quantum_2001}.
While all these approaches have had varying levels of success, there still remains a gap: very few calculations have been performed that approach the exact solution to the Schr\"odinger equation on strongly correlated systems and analyze the nature of the correlated wave function in these challenging materials.

This article is meant as a first step at understanding the relevant correlations in transition metal oxides.  
Quantum Monte Carlo(QMC) is chosen as a vehicle to do this because of two major considerations.
First, the strong dynamic correlation that is present in transition metal oxides is easily described using explicit correlation, which is efficiently evaluated using Monte Carlo techniques.
Second, and perhaps more importantly, the quantum Monte Carlo methods are able to perform calculations on extended systems efficiently, which is unique for an explicitly correlated wave function based method.
Learning what elements of the wave function are necessary for accurate treatment of transition metal oxide molecules in QMC with thus provide a valuable guide for larger molecules and extended systems.
In this article, we will explore in what ways electron correlation breaks single particle symmetry and thus discover what terms should be in an effective model of electron correlation.

\section{Symmetry and electron correlation}

Let's start witha  discussion of how one particle symmetries are broken with electron correlation.
Suppose the one-particle Hamiltonian $H_{1b}$ has a symmetry such that it commutes with the single particle symmetry operators $A_i$ and $B_i$, where $i$ refers to the single-particle electron number.
We will consider without loss of generality only one symmetry operator A.
Then the eigenstates of $H_{1b}$ can be labeled based on their symmetries as follows
\begin{align}
|s\rangle=|E;a_1,a_2,a_3,a_4,\ldots\rangle,
\end{align}
where $E$ is the eigenvalue of $H_1b$, and $\{a\}$ is the set of one-electron eigenvalues of $A$.

Now suppose that we add to the Hamiltonian a two-particle effective interaction $H_{2b}$ so that the full Hamiltonian
is $H_{1b}+H_{2b}$.  
Further suppose that $H_{2b}$ commutes with the single-particle operator $A_i$.
Then for two eigenstates of $H_{1b}$ $|s_i\rangle$ and $|s_j\rangle$, we can label them as 
\begin{align}
|s_i\rangle=|E_i,\{a\}_i\rangle,\\
|s_j\rangle=|E_j,\{a\}_j\rangle, \notag.
\end{align}
Then,
\begin{align}
\langle s_i | H_{2b} | s_j \rangle =0  \label{eqn:ondiagonal}, \\
\text{if } \{a\}_i \neq \{a\}_j 
\end{align}
That is, $H_{2b}$ does not change the block-diagonal nature of $H_{1b}$ if it commutes with $A_i$.  
This can occur when $H_{2b}$ is diagonal in any basis that preserves the symmetry of $H_{1b}$, among other cases.



Let's now contrast a realistic real-space interacting Hamiltonian with the above $H_{2b}$.
For electrons, the ab-initio Hamiltonian is $H_{1b}+\sum_{ij} 1/r_{ij}$ in atomic units.  
The $1/r_{ij}$ term conserves the overall symmetry of the system, but does not conserve particle-by-particle symmetries.  
More explicitly, the eigenstates can be labeled with the total symmetry of the state $\sum_i A_i$, where the sum is over all electrons, but not the individual $A_i$'s.

For example, if the system has cylindrical symmetry, the total angular momentum is a good quantum number for the interacting Hamiltonian, but the angular momentum of a particular electron is not.
If one attempts to emulate the effect of the $1/r_{ij}$ term by using an interaction term that conserves the single particle angular momentum, then one is enforcing the symmetry on a particle-by-particle basis.
Because conserving the one-particle symmetry aids in solving the model system, many commonly used models for electron correlation obey Eqn~\ref{eqn:ondiagonal} for at least some single particle symmetries.

For modeling transition metal oxides, a very common effective interaction is the on-site d-orbital $\hat{H}=\sum_i U \hat{n}_i^\uparrow \hat{n}_i^\downarrow$, where $\hat{n}_i$ is the number operator on an atomic-like d-orbital.  
This interaction explicitly does not allow eigenstates that are mixtures of single particle rotational states.
For example, if the nominal ground state of the transition metal monoxide MnO is (3d$^5$2p$^3$, 2p$^3$), where the states before/after the commas indicate spin up/down, then superimposing the configuration (3d$^4$4s$^1$,2p$^3$,2p$^2$3d$^1$), which involves a double electron hopping, is not allowed.  
We shall see from accurate calculations of the electronic structure of first-principles systems that this superposition is critical when considering the first principles Hamiltonian.

\section{Method}

To obtain accurate first-principles results, we use variational quantum Monte Carlo (VMC) and fixed-node diffusion Monte Carlo\cite{foulkes_quantum_2001} (FN-DMC).  
Quantum Monte Carlo methods are well-described elsewhere in the literature\cite{foulkes_quantum_2001}, so they will be described very briefly here.
VMC is a straightforward implementation of the variational method using Monte Carlo evaluation of the energy expectation value.   FN-DMC simulates the imaginary time Schr\"odinger equation is simulated to obtain the lowest energy state consistent with a given nodal surface, which further improves over the variational results.

The core electrons are replaced with a pseudopotential\cite{ovcharenko_soft_2001,lee_pseudopotentials_2000}, which is treated in the locality approximation\cite{mitas_nonlocal_1991}.  
In the transition metals, the 3s and 3p electrons are considered part of the valence.
All calculations are performed using the QWalk\cite{wagner_qwalk:_2009} package.
The trial wave function is the multi Slater Jastrow(MSJ) wave function:
\begin{equation}
\Psi(\bR)=\exp(U) \sum_k c_k {\text Det}[\phi_i^{(k)} (r_j)] ,
\end{equation}
where the determinants to include are taken from a configuration interaction in singles and doubles calculation, the one-particle orbitals are taken from a hybrid B3LYP\cite{becke:5648} calculation in GAMESS\cite{gordon_advances_2005,schmidt_general_1993}, and the Jastrow factor $U$ is the one described in Ref~\cite{wagner_energetics_2007}.
The coefficients $c_k$ are energy optimized\cite{umrigar_alleviation_2007}  simultaneously with the Jastrow parameters.
A similar approach has been shown\cite{petruzielo_approaching_2012} to efficiently produce high accuracy on a benchmark set of molecules.
Enough determinants were included that the one-particle density matrices did not change upon including more determinants.

\subsection{Calculation of the reduced density matrices}
The reduced density matrices are evaluated in quantum Monte Carlo using the following integrals
for the single particle reduced density matrix (1-RDM):
\begin{equation}
\rho_{i,k}^{(1)} = \sum_a \int { \phi_k^*(r_a') \phi_i(r_a) \Psi^*(R'_a) \Psi(R) dr_a'dR}
\end{equation}
and for the two particle reduced density matrix (2-RDM):
\begin{align}
\rho_{ij,k\ell}^{(2)} = \sum_{a\neq b} \int  \phi_k^*(r_a')\phi_\ell^*(r_b') \phi_i(r_a)\phi_j(r_a) \\
\times \Psi^*(R''_{ab}) \Psi(R)  dr_a'dr_b'dR, \notag
\end{align}
where $R=(r_1,r_2,\ldots,r_N)$, $R'_a=(r_1,r_2,\ldots,r_a',\ldots,r_N)$, and $R''_{ab}=(r_1,r_2,\ldots,r_a',\ldots,r_b',\ldots,r_N)$ and 
normalization is omitted.  
These matrices can be spin resolved, resulting in two 1-RDMs for spin up and down, and three 2-RDMs for the combination of up/up, up/down, and down/down.

In VMC, these can be evaluated by sampling two additional coordinates $r_a'$ and $r_b'$ in addition to the many-electron coordinate $R$.
In this work, the additional coordinates are sampled from the distribution $f(r)=\sum_i \phi_i^2(r)$.  $R$ is drawn as usual from $\Psi^2(R)$.  
The expectation values are thus given as follows (after a some rearrangement of terms and inclusion of normalization):
\begin{equation}
\rho_{i,k}^{(1)} = \sum_a \frac{\left\langle \frac{\Psi(R')}{\Psi(R)} \frac{\phi_k^*(r_a')\phi_i(r_a)}{f(r_a')} \right\rangle}
{N_i N_k}
\end{equation}
and similarly:
\begin{equation}
\rho_{ij,k\ell}^{(2)} = \sum_{a\neq b} \frac{\left\langle \frac{\Psi(R''_{ab})}{\Psi(R)} \frac{\phi_k^*(r_a')\phi_\ell^*(r_b')\phi_i(r_a)\phi_j(r_b)}{f(r_a')f(r_b')} \right\rangle}
{N_iN_jN_kN_\ell},
\end{equation}
with
\begin{equation}
N_i=\sqrt{\left\langle  \frac{\phi_i^2(r_a')}{f(r_a')} \right\rangle}.
\end{equation}
Choosing $f(r)$ properly to sample $r_a'$ and $r_b'$ increases the efficiency substantially, as well as using symmetry to evaluate $\Psi(R'_a)$ for all $a$ for a Slater-Jastrow or multi Slater-Jastrow wave function with little work.  
The full implementation can be found in the QWalk code.
The density matrices are expressed in a basis of B3LYP one-particle orbitals, except where noted in the text.
The density matrices using the mixed estimator in DMC are indistinguishable from the VMC results when for the converged wave functions in this work, which further reinforces the accuracy of the wave functions.

The above procedure was performed for the early transition metal monoxides ScO,TiO,VO, CrO, and MnO, and the late transition metal dioxides MnO$_2$, FeO$_2$, and CoO$_2$.  
The latter set has an interesting transition from a bent to straight bond that is very sensitive to the treatment of electron correlation.

\section{Results and discussion}
\subsection{ Geometry and dipole moments}

For the transition metal monoxides, the dipole moment(Table~\ref{table:dipole_moments}) is very challenging to calculate using 
quantum Monte Carlo.  
For the accurate wave functions considered here, the agreement with experiment is much better than using a Slater-Jastrow wave function,
although it still appears quite difficult to converge the dipole moment, since the energy of a state is not very sensitive to the dipole moment.

\begin{table}
\begin{ruledtabular}
\begin{tabular}{lcccccc}
Method & ScO & TiO & VO & CrO & MnO  \\
LDA\cite{furche:044103} & 3.57 & 3.23 & 3.10 & 3.41 & --  \\
CCSD(T)\cite{baushlicher:189} & 3.91 & 3.52 & 3.60 & 3.89 & 4.99 \\ 
TPSSh\cite{furche:044103} & 3.48 & 3.43 & 3.58 & 3.97 & --  \\
RMC(SJ) &  4.61(5) & 4.11(5) & 4.64(5) & 4.76(4) & 5.3(1)  \\
DMC(MSJ) & 3.77(2) & 3.16(2) & 3.89(5) & 3.27(4) & 4.92(4) \\
Exp\cite{steimle_review} & 4.55 & 3.34(1)\cite{steimle_tio_03} & 3.355 & 3.88 & -- \\
\end{tabular}
\end{ruledtabular}
\caption{Dipole moments in Debye.  The fixed-node RMC results have been obtained with a 
single deteriminant of B3LYP orbitals.  }
\label{table:dipole_moments}
\newpage
\end{table}

In the case of the transition metal oxides, the bond angle (Table~\ref{table:bond_angles}) is predicted poorly\cite{kulik_transition-metal_2011} by most GGA methods and even hybrid methods,
so this is a stringent test of the treatment of electron correlation.  
Diffusion Monte Carlo with a Slater-Jastrow nodal surface underestimates the bond angle of FeO$_2$, and gives a flat potential energy surface for CoO$_2$ around the linear geometry.
A more accurate wave function fixes these defects and clearly agrees with the bond angles obtained in experiment.

\begin{table}
\begin{ruledtabular}
 \begin{tabular}{ccccc}
 Method & MnO$_2$ & FeO$_2$ & CoO$_2$ \\
  GGA(PBE)\cite{kulik_transition-metal_2011} & 128 & 138 & 158  \\
 GGA+U\cite{kulik_transition-metal_2011} & 180 & 180 & 180 \\
 B3LYP& 129 & 142 & 151  \\
 DMC(SJ)  & 140(5) & 140(5) & 160-180 \\
 DMC(MSJ) & 140(5) & 155(5) & 180(5) \\
 Experiment\cite{kulik_transition-metal_2011} &135(5) & 150(10) & 180 \\
 \end{tabular}
\end{ruledtabular}
\caption{Minimum energy bond angles at $r$=1.6\AA for the transition metal dioxide molecules.  The DMC methods both obtain the correct 
bond length of 1.6 \AA.}
\label{table:bond_angles}
\end{table}

\subsection{Occupation numbers of the one-particle density matrix }

\begin{figure}
\includegraphics[width=\columnwidth]{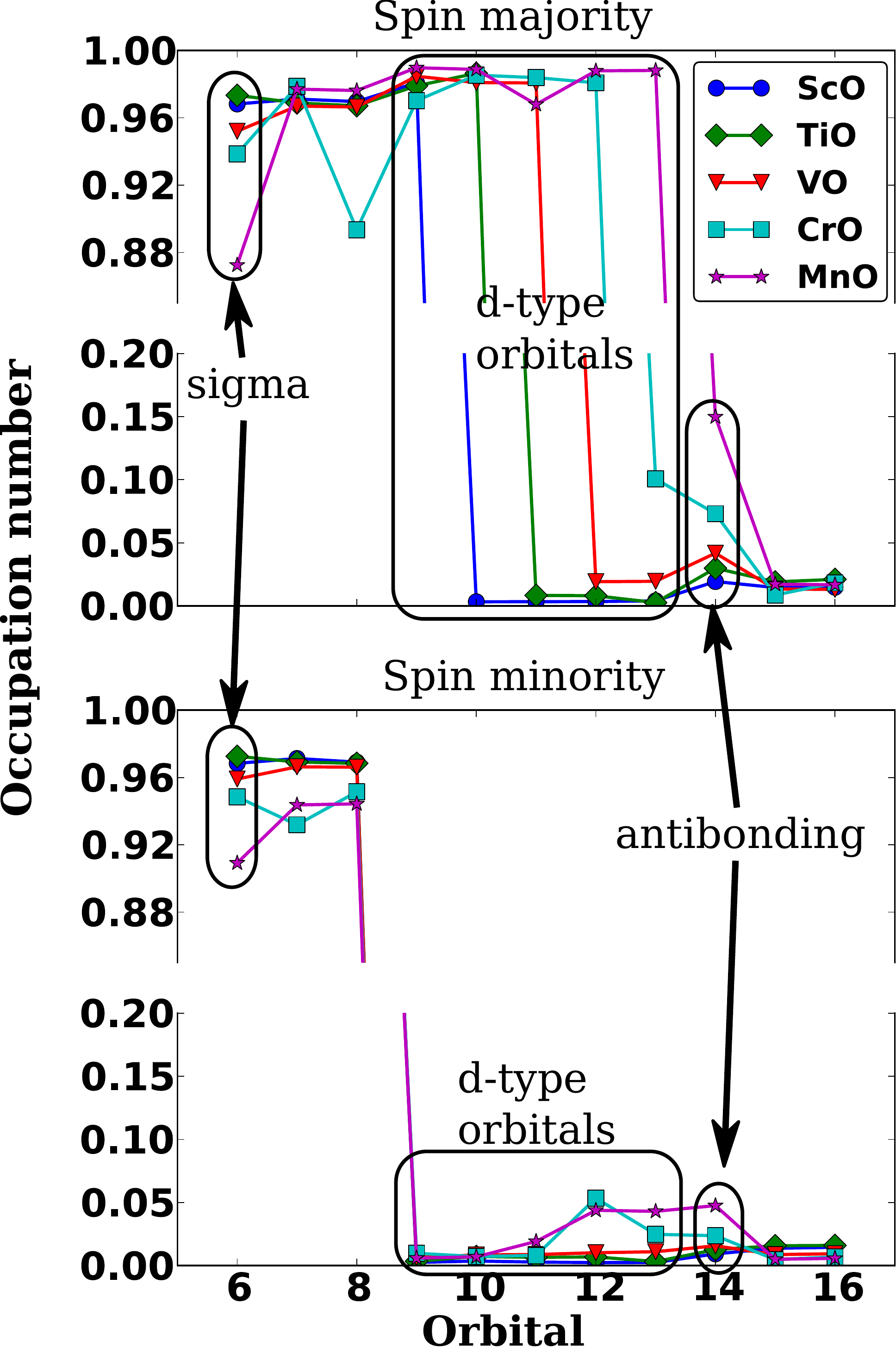}
\caption{(color online) The 1-RDM for transition metal monoxides in the B3LYP Kohn-Sham basis.  Stochastic errors are smaller than the symbol sizes.
The diagonals only are plotted, since the off-diagonal elements are very small and do not change the picture.
}
\label{fig:1rdm}
\end{figure}

In the basis of B3LYP orbitals, the single particle reduced density matrix is very accurately diagonal; there is likely little to gain in this case in orbital optimization of a single Slater determinant, beyond using B3LYP orbitals.
Thus we only report the diagonal elements (Fig~\ref{fig:1rdm}).
In the case of the monoxides, the $\sigma$-symmetry orbitals have the lowest occupation number in the nominally occupied set of states, while in the virtual space, the up electrons occupy mostly the $\sigma^*$ orbital and the down electrons occupy a number of virtual orbitals.
Interestingly, the d-like singly occupied orbitals have occupation numbers closer to 1 than the bonding-type orbitals for the spin majority, indicating that in some sense, these orbitals once occupied are not that strongly correlated.

CrO is a special case because it has a degenerate ground state in the single particle approximation.
 Correlation lifts this degeneracy by mixing the two states, which is responsible for some of the outliers in Fig~\ref{fig:1rdm}.

\subsection{Off-diagonal elements of the two particle reduced density matrix}

To analyze the breaking of one-particle symmetry by the interaction, we can turn to the two particle reduced density matrix.
Suppose that we expand the a state $|\Psi\rangle$ in terms of a basis of Slater determinants.
Since this matrix can be written in second quantized form in a basis as
\begin{equation}
\rho_{ij,k\ell}^{(2)}=\langle\Psi|c_k^\dagger c_l^\dagger c_i c_j| \Psi\rangle,
\end{equation}
one can show that if $|\Psi\rangle$'s expansion contains two Slater determinants $|s\rangle$ and $|s'\rangle$ such that 
\begin{equation}
|s'\rangle = c_k^\dagger c_l^\dagger c_i c_j |s\rangle,
\label{eqn:promotion}
\end{equation}
then $\rho_{ij,k\ell}^2$ is nonzero if $(i,j) \neq (k,\ell)$.
Conversely, if there are no Slater determinants in the expansion of $|\Psi\rangle$ connected by Eqn~\ref{eqn:promotion}, then the matrix element is zero.  
We will use this to detect the satisfaction or lack thereof in Eqn~\ref{eqn:ondiagonal} for different symmetry classes.

\begin{figure}
\includegraphics[width=\columnwidth]{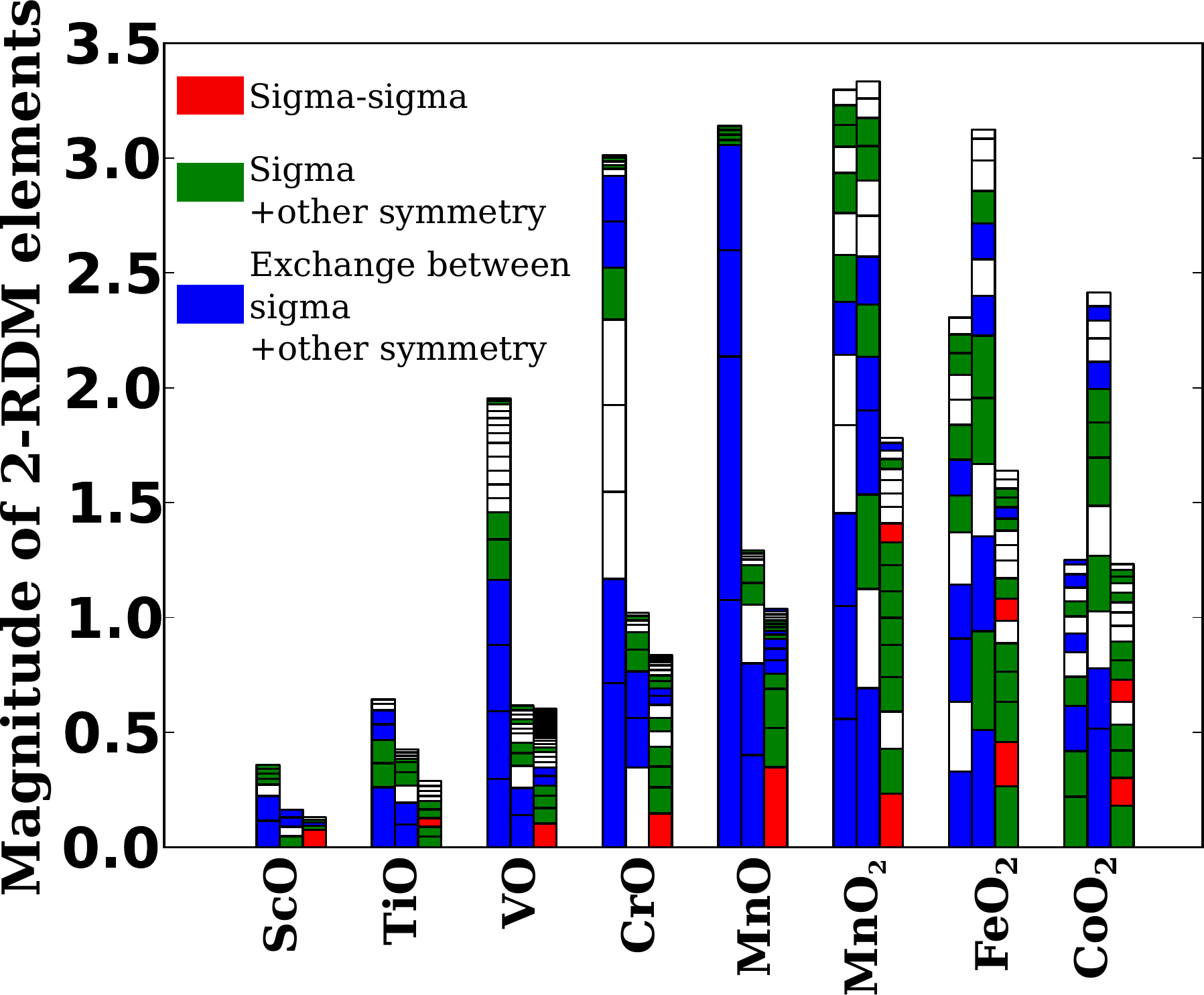}
\caption{The first column is $\uparrow\uparrow$, the second is $\downarrow\downarrow$, and the third is $\uparrow\downarrow$.
All molecules are at their equilibrium geometries, except CoO$_2$, which is set to 140$^\circ$ bond angle for easier symmetry comparison.}
\label{fig:offdiagonal_measurements}
\end{figure}

Most of the non-zero off-diagonal elements of the two-particle reduced density matrix involve the $\sigma$-bonding-like orbitals, labeled in  Fig~\ref{fig:offdiagonal_measurements}.
The only exception here is again CrO with the degenerate single-particle ground state.
For all of the systems considered here, the spin-like off-diagonal elements are larger than the spin-unlike, which is surprising--perturbation theory implies that the spin unlike correlation should be larger. 
For most of the materials, the off-diagonal elements are in the form of an exchange between the $\sigma$-like orbitals and another symmetry orbital, which breaks the one-particle rotational symmetry.
These elements are of the form of a 2-electron hopping; a typical example of which would be 
\begin{equation}
c_{\pi\uparrow}^{\dagger}c_{\sigma_2\uparrow}^{\dagger}c_{\sigma_1\uparrow}c_{\pi\uparrow},
\end{equation}
where $\sigma_1$ and $\sigma_2$ are two different $\sigma$-symmetry states and $\pi$ is a $\pi$-symmetry state.

\begin{figure}
\includegraphics[width=\columnwidth]{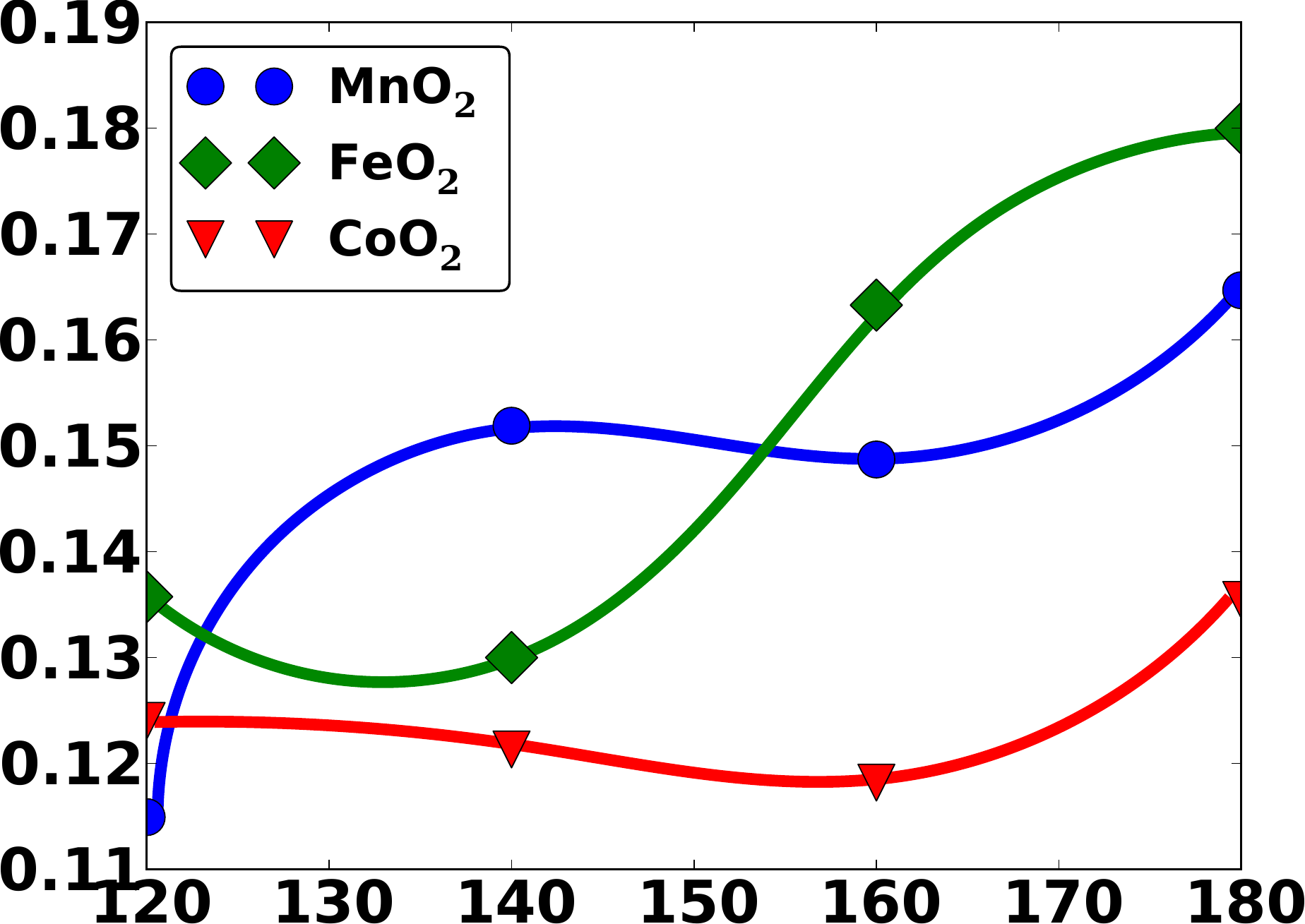} \\
\includegraphics[width=\columnwidth]{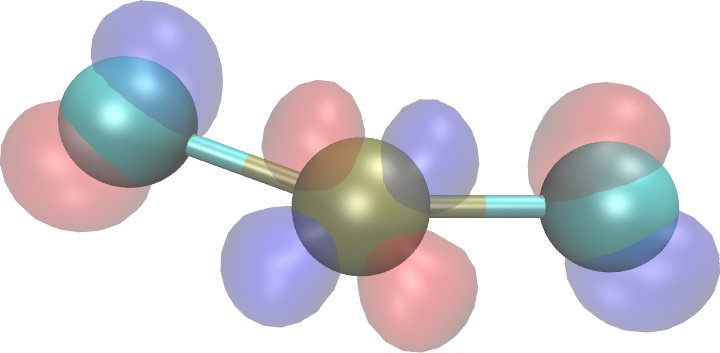}
\caption{(top) The occupation number of the 17th orbital (summed over spin) for MnO$_2$, FeO$_2$, and CoO$_2$ as a function of angle at a bond length of 1.60\AA.  The lines are guides to the eye. (bottom) The 17th orbital for MnO$_2$, with a positive isosurface in blue, and negative in red.  The other TM-O$_2$ molecules are qualitatively similar.}
\label{fig:mo2_angle}
\end{figure}

One sees a clear trend in the monoxide molecules; as the state goes from a doublet (ScO) to a sextuplet (MnO), the correlation increases monotonically.  
In the dioxides, as the state goes from a quadruplet (MnO$_2$) to a doublet (CoO$_2$), the correlation decreases monotonically.
This is likely the reason for the increase in bond angle through this series, since there are fewer empty states in CoO$_2$ in which to perform exchanges, the electrons repulse each other more as the bond angle closes, therefore tending towards a 180 degree bond.
This can be seen in the occupation number of one of the main virtual orbitals (Fig~\ref{fig:mo2_angle}) as a function of angle.

The picture emerging from the calculations can be summarized as follows.  
The $\sigma$-like bond between the transition metal and the oxygen experiences a strong dynamic interaction with other electrons.
This bond is thus most likely to be partially occupied.
In terms of virtual excitations, the most likely excitation is an electron excited from a singly occupied d-like state into a low-lying virtual orbital, and then an electron occupied from the $\sigma$-like orbital into the newly de-occupied d-like state.
Interestingly, this scenario {\em cannot} be described with an on-site Hubbard U-like term, due to the symmetry, as discussed above.
We can see the effects of this when trying to fit a low-energy model to the physics of the monoxide MnO.

\subsection{Fitting a model: MnO }

It is interesting to consider the minimal effective model that can reproduce the density matrices considered in this paper.  
As an example, the case of MnO is considered, with the valence space made from the Mn 3d and 4s states and the oxygen 2p states.  Including the 4s state is essential to reproduce the physics, since the partial occupation of this state is large. 
The full state space in this case is only 756 states, so it can be solved by exact diagonalization, and the model parameters can be optimized to reproduce the 2-RDM diagonals.  The Hamiltonian considered is
\begin{equation}
\hat{H}= \hat{E} + \hat{T}+\hat{U}_{\text{intrasite}}+ \hat{U}_{\text{2-exchange}},
\end{equation}
where $\hat{E}$ is the one-particle energy of the localized orbitals (3 parameters for 2p,3d,and 4s), $\hat{T}$ is the one-particle hopping parameter (one parameter for $\pi$-symmetry orbitals, and 3 parameters for $\sigma$-symmetry orbitals), $\hat{U}_{\text{intrasite}}$ is the Hubbard $U$ and onsite interactions that are diagonal in the local basis, and $\hat{U}_{\text{2-exchange}}$ is a 2-electron hopping term that is off-diagonal in the localized basis:
\begin{equation}
\hat{U}_{\text{2-exchange}}= \sum_{\zeta_1 \zeta_2} U_{\zeta_1 \zeta_2} c_{\zeta_1}^\dagger c_{\zeta_2}^\dagger c_{\zeta_2} c_{\zeta_1},
\end{equation}
where $\zeta$ is a spin and site index $(i,\sigma)$.
More details on the precise Hamiltonian is available in the Appendix.
The last term is critical; without it, the 2-RDM is very poorly reproduced, having RMS errors of $\pm$ 0.17 on numbers that vary between zero and one.  
With the last term, the RMS errors are reduced to $\pm$ 0.01.  
It thus appears that a 2-electron exchange term is critical to describe the correlation between the transition metal and the oxygen.
This intersite exchange term appears to be very rarely considered in theoretical descriptions of strongly correlated materials, usually entering only in an intrasite form\cite{kuroki_unconventional_2008}, although it has been noted\cite{eder_intersite_1996} that a similar term can result from downfolding a Hubbard model with intersite Coulomb interaction to a t-J model.

\section{Conclusion}

Given accurate many-body wave functions for small transition metal oxide molecules, the largest correlations break single-particle rotational symmetry.
In an effective model that reproduces the two-body physics of the MnO molecule, this effect is similar to the size of the on-site Hubbard-like interaction.
It appears that to accurately describe the electron interactions in these materials, while a Hubbard-like U term can aid in obtaining rough agreement with the true ground state, accurate agreement requires breaking the one-body rotational symmetry.
It remains to be seen whether or not these effects are more or less important as the system size grows larger.
This is under current investigation.

If it is true that the correlations presented here are generally important in transition metal oxide systems, then they may provide a guide to building more accurate trial wave functions to use in quantum Monte Carlo calculations.
The basic two-particle hopping could be described with a compact wave function of only a relatively few Slater determinants, and the optimized with powerful techniques.  
Investigations on this front are also under development.

The methods presented here are quite general, and can be applied to both solids and larger molecules.  
The 1-RDM is very inexpensive to evaluate with a proper implementation, while the 2-RDM is somewhat more expensive, but can be made to scale as the number of electrons squared if localized basis functions are used.
The key concept here is to use the reduced density matrices to carve the large Hilbert space into pieces that are more easily analyzed, and to combine this with the accurate wave functions attainable using quantum Monte Carlo techniques.
Rather than relying only on energetics to fit models, the information provided by a single calculation can inform models of the electron correlation to increase the physical realism.

The author would like to acknowledge the Taub campus cluster at the University of Illinois and XSEDE Allocation TG-DMR110100 for computational resources, and NSF DMR 12-06242 for partial funding.
He would also like to thank David Ceperley for useful discussions, Jeremy McMinis for a suggestion on fast evaluation of the reduced density matrices, and Huihuo Zheng for a thorough reading of the manuscript.

\bibliography{tm}

\section{appendix}

 \subsection{Effective multiband model for the MnO molecule}

Orbital indices:

\begin{tabular}{l|c}
0 & O 2p $\sigma$ \\
1 & Mn 3d $\sigma$  \\
2 & Mn 4s $\sigma$  \\
3 & O 2p $\pi_x$  \\
4 & Mn 3d $\pi_x$  \\
5 & O 2p $\pi_y$ \\
6 & Mn 3d $\pi_y$\\
7 & Mn 3d $\delta_1$ \\
8 & Mn 3d $\delta_2$ \\
\end{tabular}

Fitted parameters for MnO

\begin{tabular}{l|c}
$E_{2p}$ & -14.97 \\
$E_{3d}$ & -15.63 \\
$E_{4s}$ & -8.85 \\
$T_\pi$ & -4.87 \\
$T_{01}$ & -4.41 \\
$T_{02}$ & -5.1 \\
$T_{12}$ & 0.75 \\
$U_{2p}$ & -0.32 \\
$U_{3d}$ & 4.02 \\
$U_{4s}$ & 5.11 \\
$U_{2p-2p}$ & 0.27 \\
$U_{3d-3d}$ & 0.09 \\
$U_{3d-4s}$ & -1.32 \\
$U_{1001}$ & -2.65 \\
$U_{2002}$ & -2.19 \\
$U_{2112}$ & 4.96 \\
\end{tabular}

\begin{equation}
\hat{H}= \hat{E} + \hat{T}+\hat{U}_{\text{intrasite}}+ \hat{U}_{\text{2-exchange}},
\end{equation}

\begin{equation}
\hat{E}=\sum_{i \in \{0,3,5\}} E_{2p} \hat{n}_i +  \sum_{i \in \{1,4,6,7,8\}} E_{3d} \hat{n}_i  + \sum_{i \in \{2\}} E_{4s} \hat{n}_i 
\end{equation}

\begin{align}
\hat{T}=\sum_{ (i,j) \in \{(3,4),(5,6)\}} & T_\pi c_i^\dagger c_j  + T_{01} c_0^\dagger c_1 \\
&+ T_{02} c_0^\dagger c_2 +T_{12} c_1^\dagger c_2 + h.c. \notag
\end{align}

\begin{align}
\hat{U}_{\text{hubbard}}=\sum_{i \in \{0,3,5\}} U_{2p} \hat{n}_i^\downarrow \hat{n}_i^\uparrow 
  &+  \sum_{i \in \{1,4,6,7,8\}} U_{3d} \hat{n}_i^\downarrow  \hat{n}_i^\uparrow \\
  &+ \sum_{i \in \{2\}} U_{4s} \hat{n}_i^\downarrow \hat{n}_i^\uparrow \notag \\
\end{align}

\begin{align}
\hat{U}_{\text{intrasite}}=&\sum_{i,j \in \{0,3,5\}, \sigma,\sigma'} U_{2p-2p} \hat{n}_i^{\sigma'} \hat{n}_j^\sigma \\
    &+  \sum_{i,j \in \{1,4,6,7,8\},\sigma,\sigma'} U_{3d-3d} \hat{n}_i^{\sigma'}  \hat{n}_i^\sigma \\
    &+ \sum_{i \in \{2\},j \in \{1,4,6,7,8\}, \sigma, \sigma'} U_{3d-4s} \hat{n}_i^{\sigma'} \hat{n}_j^\sigma+h.c.
\end{align}

\begin{align}
 \hat{U}_{\text{2-exchange}}=U_{1001} c_1^\dagger c_0^\dagger c_0 c_1 + U_{2002} c_2^\dagger c_0^\dagger c_0 c_2 +U_{2112} c_2^\dagger c_1^\dagger c_1 c_2 
\end{align}

\end{document}